\documentclass[prl,twocolumn]{revtex4-1}
\usepackage{graphicx}
\usepackage{bm}
\usepackage{subfigure}
\usepackage{amsmath}
\usepackage[utf8]{inputenc}

\begin{document}
\title{Solitary Waves in Optical Fibers Governed by Higher Order Dispersion}

\author{V. I. Kruglov~$^1$ and J. D. Harvey~$^2$}

\affiliation{$^1$Centre for Engineering Quantum Systems,
School of Mathematics and Physics, The University of Queensland, Brisbane, QLD 4072, Australia,}
\affiliation{$^2$Department of Physics, University of Auckland, Private Bag 92019, Auckland, New Zealand.}

\begin{abstract}
An exact solitary wave solution is presented for the nonlinear Schr\"{o}dinger equation governing the propagation of pulses in optical fibers including the effects of second, third and fourth order dispersion. 
The stability of this soliton-like solution with ${\rm sech}^2$ shape is proven by the sign-definiteness of
the operator and an integral of the Sobolev type. The main criteria governing the existence of such stable localized pulses propagating in optical fibers are also formulated. A unique feature of these soliton-like optical pulses propagating in a fiber with higher order dispersion is that their parameters 
 satisfy efficient scaling relations. The main soliton solution term given by perturbation theory is also presented when absorption or gain is included in the nonlinear Schr\"{o}dinger equation.
We anticipate that this type of stable localized pulses could find practical applications in communications, slow-light devices and ultrafast lasers.
\end{abstract}

\pacs{}

\maketitle

Optical solitons governed in fibers by the nonlinear Schr\"{o}dinger equation \cite{HT,MS} could play an important role in future high-speed communication systems and ultrafast fiber lasers.
Solitary waves governed by second and fourth order dispersion only, have been studied since the 1990s \cite{B1,B2,B3,B4,D1,B5}.
It has been found that for some conditions these quartic solitons can have decaying oscillating tails \cite{B3,B4,B5}. These studies have been based on the assumption that the third order dispersion is zero which has limited the experimental observation of quartic solitons \cite{B6}.
However, the recent advent of silicon photonics has provided a way to observe and generate
quartic solitons in specially designed silicon-based waveguides \cite{C1,C2,C3,C4,C5,C6,C7,C8,C9,C10,C11}. Experimental
and numerical evidence for pure-quartic solitons and periodically modulated propagation for the higher-order quartic soliton has been reported in a recent paper \cite{AB}.   
Furthermore, photonic crystal and other types of waveguide structures have now been developed to the point where a wide range of higher order dispersion profiles can be designed and engineered.

In this paper we present an exact stationary soliton-like solution of the generalized nonlinear Schr\"{o}dinger equation (NLSE) with second, third and fourth order dispersion terms. This stable solution has a group velocity which depends on all orders of dispersion.  The soliton-like solution has been derived by a regular method which will be published elsewhere. The stability of this soliton-like solution is also demonstrated. In addition, we have also found an approximate soliton solution in the case when an absorption or gain term is included in the nonlinear Schr\"{o}dinger equation. Finally we present the main criteria for the existence of such stable solitary waves propagating in optical fibers. 

For the standard assumptions of slowly varying envelope, instantaneous nonlinear response, and no higher order nonlinearities, the generalized NLSE for the pulse envelope $\psi(z,\tau)$ has the form \cite{A1,A2,A3},
\begin{equation}
i\frac{\partial\psi}{\partial z}=\alpha\frac{\partial^2\psi}{\partial \tau^2}+i\sigma\frac{\partial^3\psi}{\partial \tau^3} 
-\epsilon\frac{\partial^4\psi}{\partial \tau^4}-\gamma |\psi|^2\psi-i\frac{\mu}{2}\psi,
\label{1}
\end{equation}
where $z$ is the longitudinal coordinate, $\tau=t-\beta_1z$ is the retarded time, and 
$\alpha=\beta_2/2$, $\sigma=\beta_3/6$, $\epsilon=\beta_4/24$, and $\gamma$ is the nonlinear parameter. The parameters 
$\beta_k=(d^k\beta/d\omega^k)_{\omega=\omega_0}$ are the k-order dispersion of the optical fiber and $\beta$ is the propagation constant. The last term in the NLSE describes absorption or amplification depending on the sign of parameter $\mu$. 

We have found the following exact solitary wave solution of Eq. (\ref{1}) for $\mu=0$:
\begin{equation}
\psi(z,\tau)={\rm u}~{\rm sech}^2[{\rm w}(\tau-\eta-{\rm v}^{-1}z)]\exp[i(\kappa z-\delta\tau+\phi)],
\label{2}
\end{equation}
where $\eta$ and $\phi$ represent the position and phase of the stable localized pulse at $z=0$.
The amplitude and inverse temporal width of the solitary wave are given by
\begin{equation}
{\rm u}=\sqrt{\frac{-3}{10\gamma\epsilon}}\left(\frac{3\sigma^2}{8\epsilon}-\alpha \right),~~~
{\rm w}=\frac{1}{4}
\sqrt{\frac{4\alpha}{5\epsilon}-\frac{3\sigma^2}{10\epsilon^2}},
\label{3}
\end{equation}
where $\alpha<0$, $\epsilon<0$ and $8\alpha\epsilon>3\sigma^2$ with $\gamma>0$.
The velocity ${\rm v}$ of the solitary wave in the retarded frame and the parameters $\delta$ and $\kappa$ are
\begin{equation}
{\rm v}=\frac{8\epsilon^2}{\sigma(\sigma^2-4\alpha\epsilon)},~~~~~~~~~~~~\delta=-\frac{\sigma}{4\epsilon},
\label{4}
\end{equation}
\begin{equation}
\kappa=-\frac{4}{25\epsilon^3}\left(\frac{3\sigma^2}{8}-\alpha\epsilon \right)^2-\frac{\sigma^2}{16\epsilon^3}\left(\frac{3\sigma^2}{16}-\alpha\epsilon \right).
\label{5}
\end{equation}
The substitution of the retarded time $\tau=t-\beta_1z$ into Eq. (\ref{2}) shows that $\delta$ and $q=\kappa+\beta_1\delta$ are the frequency and wave number shifts respectively. 
This solitary wave solution we call a soliton below for simplicity. We emphasize that this soliton does not have non-trivial free parameter. Moreover
the velocity of such solitons is fixed because the generalized  NLSE is not invariant with respect to Galilean transformations.
 Equation (\ref{3}) with $\gamma>0$ yields the next relations $\epsilon<0$ and $\alpha<3\sigma^2/8\epsilon$.
Hence the velocity is positive ${\rm v}>0$ when $\beta_3<0$, and the velocity is negative ${\rm v}<0$ when $\beta_3>0$.
In the case when $\beta_3=0$ the solution reduces to that given in \cite{B2}. 
Equation (\ref{1}) for $\mu=0$ can also be written as
\begin{equation}
i\frac{\partial\psi}{\partial z}=-\frac{\delta{\cal H}}{\delta \psi^{*}},
\label{6}
\end{equation}
where ${\cal H}$ is the Hamiltonian of the system. The stability of this soliton solution is proven by the sign-definiteness of the operator and an integral of the Sobolev type. This method  \cite{B5} yields the stability region which is the same as the region of existence of ${\rm sech}^2$ solitons: 
$\beta_2<0$, $\beta_4<0$, and $2\beta_2\beta_4>\beta_3^2$, where $\beta_3$ can be negative, positive or zero. The proof is based on the boundedness of the Hamiltonian for a fixed value of soliton energy and an explicit soliton solution given in Eq. (\ref{2}). 
The energy ${\rm E}$ of the solitons for $\mu=0$ is given by,
\begin{equation}
{\rm E}=\int_{-\infty}^{+\infty}|\psi(z,\tau)|^2d\tau=\frac{4}{\gamma\sqrt{5|\epsilon|}}\left(\frac{3\sigma^2}{8\epsilon}-\alpha\right)^{3/2}.
\label{7}
\end{equation}

Note that the energy ${\rm E}$ and other parameters of the solitons satisfy simple scaling relations if the dispersion parameters are defined in the form: $\beta_k=\beta_k^{(0)} q$ where $k=2,3,4$ and $q$ is a positive dimensionless parameter. In this case the scaling relations are
\begin{equation}
{\rm E}={\rm E}_0 q,~~{\rm u}={\rm u}_0 q^{1/2},~~{\rm v}={\rm v}_0 q^{-1},~~\kappa=\kappa_0 q,
\label{8}
\end{equation}
and also we have ${\rm w}={\rm w}_0$ and $\delta=\delta_0$.
Here ${\rm E}_0$ is given by Eq. (\ref{7}) with the change $\alpha\mapsto\alpha_0$, $\sigma\mapsto\sigma_0$ and 
$\epsilon\mapsto\epsilon_0$ where $\alpha_0=\beta_2^{(0)}/2$, $\sigma_0=\beta_3^{(0)}/6$ and $\epsilon_0=\beta_4^{(0)}/24$. 
The same change is assumed for all other relations in Eq. (\ref{8}). Thus if the parameter $q$ grows the energy ${\rm E}$ of the solitons and the absolute value of the inverse velocity $|{\rm v}|^{-1}$ grow proportional to parameter $q$. It also follows from Eq. (\ref{8}) that in this case the amplitude ${\rm u}$ of the solitons grows as $q^{1/2}$. However the width ${\rm w}^{-1}$ and the wave number shift $\delta$ do not change when the parameter $q$ grows. Note that the velocity of the solitons is given by ${\rm v}_s={\rm v}/(1+\beta_1{\rm v})$. Hence the velocity of the solitons tends to zero when $q\rightarrow \infty$ because Eq. (\ref{8}) yields the scaling relation ${\rm v}_s={\rm v}_0/(q+\beta_1{\rm v}_0)$. Nevertheless the value of the parameter $q$ is limited in optical fibers. Thus we have demonstrated that it is possible to create a new type of solitary wave propagating with reduced speed and  high energy with suitable dispersion profiles. 

We anticipate that the scaling feature of ${\rm sech}^2$ solitons can find various practical applications. As an example,
tunable all-optical delay systems that dynamically manipulate the group velocity of light have received a great deal of attention for optical information processing applications, such as data buffering and synchronization. Various slow-light devices have been explored as potential realizations of a practical delay system \cite{S1,S2,S3,S4,S5}.
It is efficient to reduce the number of parameters of the NLSE using appropriate dimensionless variables. Without loss of generality
we can define the next new variables,
\begin{equation}
\psi(z,\tau)={\rm Q}{\rm U}(\zeta,\xi),~~~{\rm Q}=\frac{|\alpha|}{\sqrt{\gamma|\epsilon|}},
\label{9}
\end{equation}
where $\zeta=z/l$ and $\xi=\tau/\tau_0$. We also define here the length $l=|\epsilon|/\alpha^2$ and time $\tau_0=\sqrt{\epsilon/\alpha}$ with $\alpha<0$ and $\epsilon<0$. In this case Eq. (\ref{1}) has the dimensionless form,
\begin{equation}
i\frac{\partial{\rm U}}{\partial \zeta}=-\frac{\partial^2{\rm U}}{\partial \xi^2}+i\lambda\frac{\partial^3{\rm U}}{\partial \xi^3} 
+\frac{\partial^4{\rm U}}{\partial \xi^4}-|{\rm U}|^2{\rm U}-i\frac{\Gamma}{2}{\rm U},
\label{10}
\end{equation}
where $\lambda=\sigma/\sqrt{\alpha\epsilon}$ and $\Gamma=\mu l=\mu|\epsilon|/\alpha^2$
are two dimensionless parameters. We emphasise that $\lambda$
does not depend on the parameter $q$ when we consider the scaling relations given in Eq. (\ref{8}). However the dimensionless parameter $\Gamma=\Gamma_0 q^{-1}$ tends to zero when $q\rightarrow \infty$.

In the case when $\Gamma=0$ the soliton solution of Eq. (\ref{10}) depends on a single fiber parameter $\lambda$ and has the form,
\begin{equation}
{\rm U}(\zeta,\xi)=u_\lambda{\rm sech}^2[w_\lambda(\xi-\xi_0-v_\lambda^{-1}\zeta)]\exp[i\Phi(\zeta,\xi)],
\label{11}
\end{equation}
where $\lambda^2<8/3$ and $\xi_0$ is the position of the soliton at $\zeta=0$. The dimensionless inverse velocity and the phase of the soliton are
\begin{equation}
v_\lambda^{-1}=\lambda^3/8-\lambda/2,~~~\Phi(\zeta,\xi)=k_\lambda\zeta-d_\lambda\xi+\phi.
\label{12}
\end{equation}
The amplitude $u_\lambda$ and inverse width $w_\lambda$ of soliton are
\begin{equation}
u_\lambda=\sqrt{\frac{3}{10}}\left(1-\frac{3\lambda^2}{8}\right),~~~w_\lambda=\sqrt{\frac{1}{20}\left(1-\frac{3\lambda^2}{8}\right)}.
\label{13}
\end{equation}
Thus the amplitude $u_\lambda$ and width $w_\lambda^{-1}$ of soliton are related by $u_\lambda w_\lambda^{-2}=2\sqrt{30}$.
The functions $k_\lambda$ and $d_\lambda$ connected to the wave number and frequency shifts of the soliton are given by
\begin{equation}
k_\lambda=\frac{4}{25}\left(\frac{3\lambda^2}{8}-1\right)^2+\frac{\lambda^2}{16}\left(\frac{3\lambda^2}{16}-1\right),~~~d_\lambda=\frac{\lambda}{4}.
\label{14}
\end{equation}
In Fig. 1 we show the shape $|{\rm U}|$ of solitons in Eq. (\ref{11}) for different values of dimensionless parameter 
$\lambda$: $\lambda_0=0$, $\lambda_1=\pm 0.6$, $\lambda_2=\pm 0.9$, $\lambda_3=\pm 1.2$ and $\lambda_4=\pm 1.5$. 
We also plot in  Fig. 2 the inverse velocity $v_\lambda^{-1}$ and the inverse temporal width $w_\lambda$ of the solitons for the region 
$|\lambda|< \sqrt{8/3}$.

\begin{figure}
\includegraphics[width=9cm,trim=0mm 16mm 0mm 0mm]{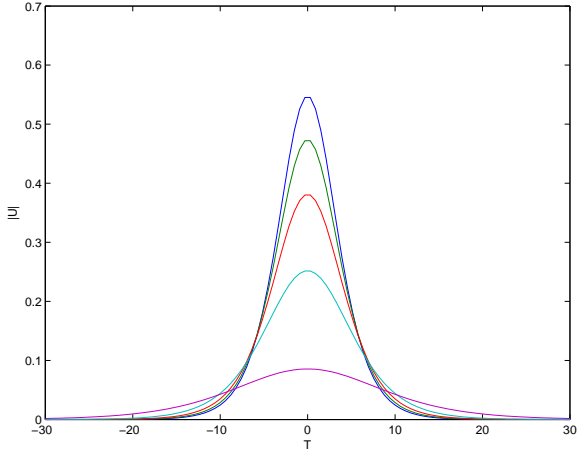}
\caption{Shape $|{\rm U}|=u_\lambda{\rm sech}^2(w_\lambda{\rm T})$ of soliton for $\lambda=\lambda_n$:
$\lambda_0=0$, $\lambda_1=\pm 0.6$, $\lambda_2=\pm 0.9$, $\lambda_3=\pm 1.2$ and $\lambda_4=\pm 1.5$. Peak of amplitude monotonically decreases for increasing $|\lambda_n|$.} 
\label{FIG.1.}
\end{figure}

\begin{figure}
\includegraphics[width=9cm,trim=0mm 16mm 0mm 0mm]{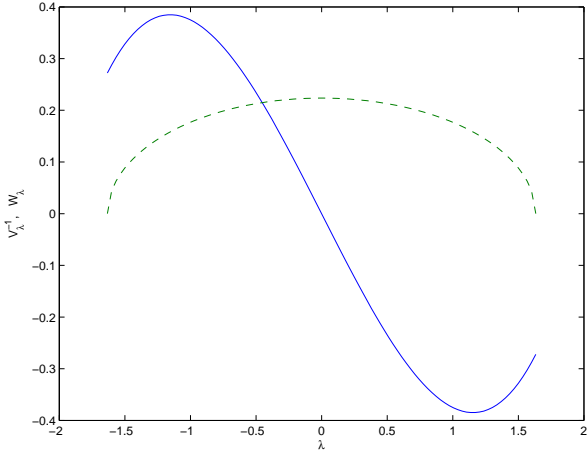}
\caption{Inverse velocity $v_\lambda^{-1}$ (solid line) and  inverse temporal width $w_\lambda$ (dashed line) of  soliton for
 $-\sqrt{8/3}<\lambda<\sqrt{8/3}$.}
\label{FIG.2.}
\end{figure}

\begin{figure}
\includegraphics[width=9cm,trim=0mm 16mm 0mm 0mm]{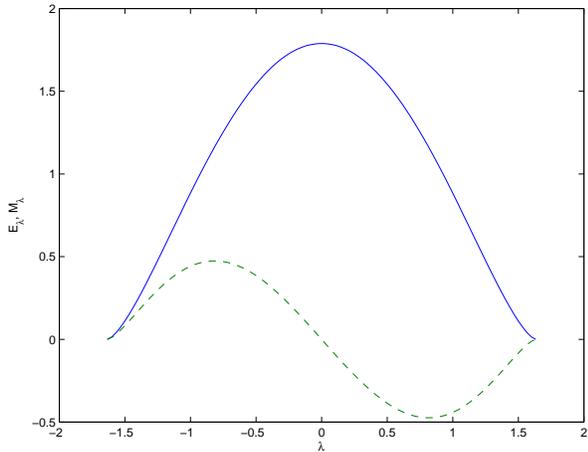}
\caption{Energy $E_\lambda$ (solid line) and  momentum  $M_\lambda$ (dashed line) of soliton for interval $-\sqrt{8/3}<\lambda<\sqrt{8/3}$.}
\label{FIG.3.}
\end{figure}
The dimensionless energy of the soliton is the integral of motion when $\Gamma=0$. In this case we have 
\begin{equation}
E_\lambda=\int_{-\infty}^{+\infty}|{\rm U}|^2d\xi=\frac{4}{\sqrt{5}}\left(1-\frac{3\lambda^2}{8}\right)^{3/2}.
\label{15}
\end{equation}
Another integral of motion when $\Gamma=0$ is the momentum. The dimensionless momentum of the soliton is 
 \begin{eqnarray}
M_\lambda=\int_{-\infty}^{+\infty}i\left({\rm U}\frac{\partial{\rm U}^{*}}{\partial \xi}-{\rm U}^{*}\frac{\partial{\rm U}}{\partial \xi}   \right)d\xi
\nonumber\\ \noalign{\vskip3pt}       =-\frac{2\lambda}{\sqrt{5}}\left(1-\frac{3\lambda^2}{8}\right)^{3/2}.                                                                                                                                 
\label{16}
\end{eqnarray}
Hence we have the relation $M_\lambda=-(\lambda/2) E_\lambda$. In Fig. 3 the energy $E_\lambda$ and momentum  $M_\lambda$ of solitons are plotted over the region $|\lambda|< \sqrt{8/3}$.

We consider below a more general case when the last term in Eq. (\ref{10}) is taken into account. First we transform Eq. (\ref{10})
to new variables given by 
\begin{equation}
{\rm U}(\zeta,\xi)=e^{-\Gamma\zeta}{\rm V}(Z,T),
\label{17}
\end{equation}
where $Z=(1-e^{-2\Gamma\zeta})/2\Gamma$ and $T=T_0+e^{-\Gamma\zeta}(\xi-\xi_0)$.
In this case the equation for the function ${\rm V}(Z,T)$ is
\begin{eqnarray}
i\frac{\partial{\rm V}}{\partial Z}+\frac{\partial^2{\rm V}}{\partial T^2}-i\lambda \sqrt{1-2\Gamma Z}\frac{\partial^3{\rm V}}{\partial T^3} 
-(1-2\Gamma Z)\frac{\partial^4{\rm V}}{\partial T^4}
\nonumber\\ \noalign{\vskip3pt}    +|{\rm V}|^2{\rm V}=\frac{i\Gamma}{1-2\Gamma Z}\left[\frac{{\rm V}}{2}+(T-T_0)\frac{\partial{\rm V}}{\partial T}\right].~~~~
\label{18}
\end{eqnarray}
We assume below that two conditions are satisfied: $|\Gamma|\ll 1-2\Gamma Z$ and $2|\Gamma Z|\ll 1$, which are equivalent to the next relations $|\Gamma|\ll e^{-2\Gamma\zeta}$ and $|1-e^{-2\Gamma\zeta}|\ll 1$. In this case the soliton solution of Eq. (\ref{18}) is given by Eq. (\ref{11})
with the change $U\mapsto V$, $\zeta\mapsto Z$ and $\xi-\xi_0\mapsto T-T_0$. The substitution of the resultant function ${\rm V}(Z,T)$ into Eq. (\ref{17}) yields the main term of the perturbation theory as 
\begin{eqnarray}
{\rm U}(\zeta,\xi)= u_\lambda e^{-\Gamma\zeta}{\rm sech}^2\{w_\lambda[e^{-\Gamma\zeta}(\xi-\xi_0)-v_\lambda^{-1}f(\zeta)]\}
\nonumber\\ \noalign{\vskip3pt}   \times\exp[ik_\lambda f(\zeta)-id_\lambda e^{-\Gamma\zeta}(\xi-\xi_0)+i(\phi-d_\lambda\xi_0)] ,~~~~~
\label{19}
\end{eqnarray}
where $f(\zeta)=(1-e^{-2\Gamma\zeta})/2\Gamma$. It is assumed here that two  
conditions $|\Gamma|\ll 1$ and $2|\Gamma\zeta|\ll 1$ are satisfied.
This approach is based on an extension of the perturbation method developed in \cite{K1,K2}.
The transformation of the solution in Eq. (\ref{19}) to the function $\psi(z,\tau)$ by Eq. (\ref{9}) yields
the approximate soliton solution of the generalized NLSE as
\begin{eqnarray}
\psi(z,\tau)={\rm u} e^{-\mu z}{\rm sech}^2\{{\rm w}e^{-\mu z}[\tau-\eta-{\rm v}^{-1}\mu^{-1}{\rm sh}(\mu z)]\}
\nonumber\\ \noalign{\vskip3pt}   \times \exp[i\kappa F(z)-i\delta e^{-\mu z}(\tau-\eta)+i(\phi-\delta\eta)],~~~~~~
\label{20}
\end{eqnarray}
where $F(z)=(1-e^{-2\mu z})/2\mu$. It is assumed here that two conditions are satisfied: $|\mu\epsilon|/\alpha^2\ll 1$ and $2|\mu z|\ll 1$. This equation for different signs of the coefficient $\mu$ describes decay or amplification of solitons. 
It follows from Eq. (\ref{20}) that the initial pulse at $z=0$ is given by the soliton solution in Eq. (\ref{2}). 
Note that in the limit when $\mu\rightarrow 0$ the solution in Eq. (\ref{20}) tends to an exact soliton solution given in Eq. (\ref{2}). It also follows from Eq. (\ref{20}) that the velocity of the peak soliton amplitude $|\psi(z,\tau)|$ in 
the retarded frame is ${\rm v}(z)={\rm v}~{\rm sech}(\mu z)$. 
Equation (\ref{1}) leads to differential equation for the soliton energy ${\rm E}(z)$ as 
 \begin{equation}
\frac{d{\rm E}(z)}{dz}=-\mu {\rm E}(z),~~~{\rm E}(z)=\int_{-\infty}^{+\infty}|\psi(z,\tau)|^2d\tau.
\label{21}
\end{equation}
This exact equation has the solution ${\rm E}(z)={\rm E}e^{-\mu z}$ where the energy ${\rm E}$ is given in Eq. (\ref{7}). 
It is worth noting that the approximate solution given in Eq. (\ref{20}) leads to the same soliton energy ${\rm E}(z)$ as the exact Eq. (\ref{21}). The computations show that the solution in Eq. (\ref{20}) provides a good accuracy when $|\mu\epsilon|/\alpha^2\ll 1$ and $|\mu z|\leq 0.1$. This first inequality can be satisfied when we consider the scaling relations in Eq. (\ref{8}) because in this case $|\mu\epsilon|/\alpha^2\propto q^{-1}$. 

In the case when the velocity of the solitons ${\rm v}_s=({\rm v}^{-1}+\beta_1)^{-1}$ is negative it is useful to change the coordinate system to the inverse direction. It can be shown by Eq. (\ref{1}) that such a transformation in soliton solutions $\psi(z,\tau)$ is given by $\psi\mapsto \psi^{*}$, $z\mapsto -z$, $\beta_1\mapsto -\beta_1$, $\beta_3\mapsto -\beta_3$ and $\mu\mapsto -\mu$. The solutions in Eqs. (\ref{2}) and (\ref{20}) are invariant to this transformation because the phase $\phi$ is arbitrary. However the velocity ${\rm v}$ in the retarded frame and the parameter $\beta_1$ change sign after this transformation and then the velocity of the solitons 
${\rm v}_s$ becomes positive.

Note that we have neglected in the generalized NLSE the Raman and higher order nonlinear effects which lead to the next necessary condition ${\rm w}^{-1}>\tau_c$ for the pulse width. Hence the width of solitons is restricted by some characteristic time $\tau_c$ depending on the fiber parameters. 
Moreover Eq. (\ref{13}) leads to the relation $\lambda^2<8/3$ which is a necessary condition for the existence of the soliton solution.
These two criteria for the existence of ${\rm sech}^2$ solitons can be written as
\begin{equation}
2\beta_2\beta_4>\beta_3^2,~~~~~|\beta_4|(2\beta_2\beta_4-\beta_3^2)^{-1/2}>\sqrt{0.3}\tau_c,
\label{22}
\end{equation}
 where $\beta_2<0$ and $\beta_4<0$. Note that the dispersion parameters of silicon-based structures satisfy the criteria in Eq. (\ref{22}) for appropriate geometry and materials of the structures. We also emphasize that in the case when $\beta_3<0$ the velocity ${\rm v}_s$ of the solitons is positive and the next inequality ${\rm v}_s<\beta_1^{-1}$ is satisfied. In the limiting case when $\beta_3=0$ we have the relation ${\rm v}_s=\beta_1^{-1}$ where $\beta_1^{-1}$ is the group velocity.  
 
 In summary, we have found an exact solitary wave solution of the nonlinear Schr\"{o}dinger equation including the effects of second, third and fourth order dispersion. It is shown that these soliton-like pulses with ${\rm sech}^2$ shape are stable. In the case when the absorption
or gain term is included in the NLSE the main soliton solution term given by perturbation theory has also been found, together with the main criteria governing the existence of such stable localized pulses propagating in optical fibers. Furthermore, the derived scaling relations 
show that the group velocity of  ${\rm sech}^2$ solitons can be significantly reduced for appropriate parameters of the waveguide which may find application in developing slow-light systems.

The support of the Dodd-Walls Centre for Photonic and Quantum Technologies is gratefully acknowledged.


\begin{thebibliography}{99}






\bibitem{HT} A. Hasegava and F. Tappert, Appl. Phys. Lett. {\bf 23}, 142 (1973).

\bibitem{MS} L. F. Mollenauer, R. H. Stolen and J. P. Gordon, Phys. Rev. Lett. {\bf 45}, 1095 (1980).

\bibitem{B1} A. H\"{o}\"{o}k and M. Karlsson, Opt. Lett. {\bf 18}, 1388 (1993).

\bibitem{B2} M. Karlsson and A. H\"{o}\"{o}k, Opt. Comm. {\bf 104}, 303 (1994).

\bibitem{B3} N. N. Akhmediev, A. V. Buryak, and M. Karlsson, Opt. Comm. {\bf 110}, 540 (1994).

\bibitem{B4} N. N. Akhmediev, A. V. Buryak,  Opt. Comm. {\bf 121}, 109 (1995).

\bibitem{D1} N. Akhmediev and M. Karlsson, Phys. Rev. A {\bf 51}, 2602 (1995).

\bibitem{B5} V. E. Zakharov and E. A. Kuznetsov, J. Exp. Theor. Phys. {\bf 86}, 1035 (1998).

\bibitem{B6} S. Roy and F. Biancalana, Phys. Rev. A {\bf 87}, 025801 (2013).

\bibitem{C1} {\it Silicon Photonics}, edited by L. Pavesi and D. J. Lockwood (Springer, New York, 2004).

\bibitem{C2} B. Jalali, J. Lightwave Technol. {\bf 24}, 4600 (2006).

\bibitem{C3} M. Lipson, Nanotechnol. {\bf 15}, S622 (2004).

\bibitem{C4} Q. Lin, O. J. Painter, and G. P. Agrawal, Opt. Express {\bf 15}, 16604 (2007).

\bibitem{C5} A. D. Bristow, N. Rotenberg, and H. M. van Driel, Appl. Phys. Lett. {\bf 90}, 191104 (2007).

\bibitem{C6} X. Liu {\it et al.}, Opt. Express {\bf 19}, 7778 (2011).

\bibitem{C7} L. Zhang {\it et al.}, Opt. Express {\bf 18}, 20529 (2010).

\bibitem{C8} L. Zhang {\it et al.}, Opt. Express {\bf 20}, 1685 (2012).

\bibitem{C9} Q. Lin {\it et al.}, Opt. Express {\bf 14}, 4786 (2006).

\bibitem{C10} B. Jalali {\it et al.}, IEEE J. Sel. Top. Quant. Electron. {\bf 12}, 1618 (2006).

\bibitem{C11} V. Raghunathan {\it et al.}, Opt. Express {\bf 15}, 14355 (2007).

\bibitem{AB} A. Blanco-Redondo {\it et al.}, Nature Comm. {\bf 7}, 10427 (2016).

\bibitem{A1} K. J. Blow and D. Wood, IEEE J. Quant. Electron. {\bf 25}, 2665 (1989).

\bibitem{A2} E. Golovchenko and A. N. Pilipetskii, J. Opt. Soc. Am. B {\bf 11}, 92 (1994).

\bibitem{A3} S. B. Cavalcanti, J. C. Cressoni, H. R. da Cruz and A. S. Gouveia-Neto, Phys. Rev. A {\bf 43}, 6162 (1991).

\bibitem{S1} L. V. Hau, S. E. Harris, Z. Dutton and C. H. Behroozi, Nature {\bf 397}, 594 (1999).

\bibitem{S2} F. Morichetti, A. Melloni, C. Ferrati and M. Martinelli, Opt. Express {\bf 16}, 8395 (2008).

\bibitem{S3} A. Melloni, F. Morichetti, C. Ferrati and M. Martinelli, Opt. Lett. {\bf 33}, 2389 (2008).

\bibitem{S4} F. Xia, L. Sekaric and Y. Vlasov, Nature Photon. {\bf 1}, 65 (2007).

\bibitem{S5} V. Govindan and S. Blair, J. Opt. Soc. Am. B {\bf 25}, C23 (2008).

\bibitem{K1} Y. Kodama, J. Phys. Soc. Jap. {\bf 45}, 311 (1978).

\bibitem{K2} A. Hasegawa and Y. Kodama, Proc. IEEE {\bf 69}, 1145 (1981).


\end{thebibliography}
\end{document}